\documentclass[11pt]{article}

\usepackage{amsfonts}
\usepackage{epsf}
\usepackage{graphicx}
\usepackage{psfrag}
\usepackage{amssymb}
\usepackage{color}
\usepackage{amsmath}
\usepackage{chemarr}

\usepackage{xcolor}

\newcommand{\rme}{\mathrm{e}}
\newcommand{\rmi}{\mathrm{i}}
\newcommand{\rmd}{\mathrm{d}}

\newcommand{\bfr}{\mathbf{r}}

\renewcommand{\qquad}{\hspace*{25pt}}

\begin{document}

\begin{center}
{\Large\bf Bose--Einstein Condensate Dark Matter That Involves Composites}\\

\vspace*{2.5mm}
{A.M. Gavrilik\footnote{e-mail: omgavr@bitp.kiev.ua} and A.V. Nazarenko}\\

\vspace*{1.5mm}
{\small Bogolyubov Institute for Theoretical Physics of NAS of Ukraine\\
14b, Metrolohichna Str., Kyiv 03143, Ukraine}
\end{center}

\abstract{By improving the Bose--Einstein condensate model of dark matter
through the repulsive three-particle interaction to better reproduce
observables such as rotation curves, both different thermodynamic phases
and few-particle correlations are revealed. Using the numerically found
solutions of the Gross--Pitaevskii equation for averaging the products of
local densities and for calculating thermodynamic functions at zero temperature,
it is shown that the few-particle correlations imply a first-order phase
transition and are reduced to the product of single-particle averages
with a simultaneous increase in pressure, density, and quantum fluctuations.
Under given conditions, dark matter exhibits rather the properties
of an ideal gas with an effective temperature determined by quantum
fluctuations. Characteristics of oscillations between bound and unbound
states of three particles are estimated within a simple random walk approach
to qualitatively models the instability of particle complexes.
On the other hand, the density-dependent conditions for the
formation of composites are analyzed using chemical kinetics
without specifying the bonds formed. The obtain results can be extended
to the models of multicomponent dark matter consisting of composites
formed by particles with a large scattering length.}

{{{\em Keywords:} dark matter; halo core; Bose--Einstein condensate; two-
and three-particle self-interactions; two-phase structure;
first-order phase transition; composites; random walk; chemical kinetics}

\section{Introduction}

The current observable restrictions on the properties of dark matter still admit
the construction of theoretical models with different 
constituents~\cite{A18,axion4,CM16,SNE16,Chav2018,Chav2021,SN19,Bel14,CKS16,GKV19,HBG00,Lee17,SCB20,MGUL00,
muBose1,muBose2,Sin,Bohmer,FMT08,Harko2011,Matos,Naz20,GKN20,GN21,Urena,Ferreira}.
Adopting the concept of dark matter consisting of light bosons in a condensate
state~\cite{muBose1,muBose2,Sin,Bohmer,FMT08,Harko2011,Matos,Naz20,GKN20,GN21,Urena,Ferreira},
we might also borrow the known properties of atomic Bose--Einstein condensates
in gaseous or liquid form~\cite{LL9,KDS99,Leggett}, reproduced in the laboratory. 

In the case of weakly
interacting structureless bosons with or without specification of their sort,
the description of the macroscopic properties of the dark matter halo of galaxies
is usually based on the Gross--Pitaevskii equation under the imposed conditions
of diluteness and a large healing length. Indeed, such models for the halo of
dwarf galaxies give a good qualitative description with a very short (pairwise)
scattering length of ultralight bosons~\cite{Chav2018,Harko2011,GN21}. It
turns out that attempts to achieve better agreement between theoretical
estimates and astronomical observables by taking into account additional
terms of few-particle interactions in the Gross--Pitaevskii equation result in
the existence of various thermodynamic phases of dark 
matter~\cite{Chav2018,GKN20,GN21}, as well as nontrivial correlations
associated with few-particle complexes (composites).

On the other hand, there are experimental data on the formation of Bose condensate
droplets~\cite{Cabr17,Ro07, Bulgac02} from a gas of alkali atoms with a long scattering
length, which is accompanied by the processes of particle recombination, three-particle
losses, and the formation of molecules that are technically suppressed in the laboratory.
This regime, associated with the unitary limit of infinite scattering length, implies
a strong interaction between particles leading to the structural transformations. Much
attention is paid to its study in the literature~\cite{BrN99,BrH06,Mus21} in connection
with experimental exploration of the Bose condensate. Theoretically, similar processes
like noted above cannot in principal be ruled out in dark matter, which could
eventually be a (multicomponent) mixture of different composites with a resulting short
scattering length, as required by current observations. In addition, they could take
place under the extreme conditions in the early Universe~\cite{BKL18}.
Thus, the assumptions proposed give rise to many implications, and we would like
to dwell on a few. It is worth to note that dark matter, made of
several constituents, is already being actively studied~\cite{PG19,Matos21}. 

Summing up, we outline the tasks of this paper, which is devoted to
the analysis of density-dependent conditions for the formation of composites
using rather simple models.

In Section~2, considering a model based on the Gross--Pitaevskii
equation that takes into account two- and three-particle interactions, we
try to obtain the admissibility conditions for the occurrence of particle complexes
by estimating few-particle correlators over a relatively wide range of density,
pressure, and quantum fluctuations at zero temperature, i.e., quantities that
affect the transfer momentum in an inhomogeneous one-component system of bosons.
In Section~3, we describe the time-oscillatory instability
of two- and three-particle states that arises in a dense medium using a simple
random walk model~\cite{Kempe}. Various scenarios for the formation of composites
from three particles are considered in Section~4 within the framework of chemical
kinetics~\cite{Zam} without clarifying the nature of the bonds. We operate only
with the scattering length and the density of the components as the most universal
characteristics, in order to extract the necessary conditions for the existence of those.
We end with the Discussion where we present our conclusions on the obtained results. 

\section{Self-Gravitating Bose--Einstein Condensate: The Role of Few-Particle Interactions}

We start with an analysis of the equilibrium properties of halo dark matter, described
by the Gross--Pitaevskii equation, where both two- and three-particle repulsive interactions
are taken into account. As established in \cite{GN21}, such a model reveals two-phase
structure, represented by dilute and denser (liquidlike) states, and the first-order
phase transition. It is intended to describe the resulting state of dark matter after
a long-term evolution under certain conditions, although only some of the physical
options provided by this model can be justified. After many attempts of description,
with their special features, made in the literature \cite{Harko2011,Chav2018,GKN20},
here we can expect and generally conclude that the resulting constituents of dark matter
possess a vanishingly small (pair) $s$-scattering length $a$. This does not contradict
the development of models describing particles with a longer scattering length, because
their composites (with a short scattering length $a$) are allowed to exist in
this ``final'' stage. 

So, we first refer to our exploration of the model elaborated in \cite{GN21}, where
the Bose--Einstein condensate (BEC) is described by a real function
$\psi(r)$ of the radial variable $r=|\bfr|$ in a ball $B=\{\bfr\in\mathbb{R}^3|\, |\bfr |\leq R\}$.
The main equation of our model is generated by the functional $\Gamma$, depended on a chemical
potential $\mu$, and is supplemented by the Poisson equation for gravitational field $4\pi GmV(r)$:
\begin{eqnarray}
\Gamma&=&4\pi\int_0^R\left[\frac{\hbar^2}{2m}(\partial_r\psi(r))^2-\mu\psi^2(r)
+4\pi Gm^2\psi^2(r)V(r)
\right.
\nonumber\\
&&\left.+\frac{U_2}{2}\psi^4(r)
+\frac{U_3}{3}\psi^6(r)\right]\,r^2\,\rmd r,\label{G1}\\
\Delta_r V(r)&=&\psi^2(r).
\end{eqnarray}

Let two- and three-particle interactions be characterized by the quantities:

\begin{equation}
U_2=\frac{4\pi\hbar^2a}{m}, \quad U_3=gU_2\lambda^3,
\quad \lambda=\frac{\hbar}{mc},
\end{equation}
where $a>0$ is the $s$-scattering length of pairwise interaction;
$\lambda$ is the Compton wave length of the particle with mass $m$;
$g$ is a some dimensionless parameter introduced for convenience;
$\Delta_r$ and $\Delta^{-1}_r$ are the radial part of the Laplace
operator and its inverse.

Note that the relativistic nature of three-particle interaction, because of
dependence of $\lambda$ and $U_3$ on speed of light $c$, is pointed out in \cite{Chav2018}.

Applicability of the local form of interaction requires that

\begin{equation}\label{naaa}
n_0a^3<1, \quad n_0\lambda^3<1,
\end{equation}
where $n_0$ is the maximal particle density in the system.

Vanishing of variation $\delta\Gamma$ with respect to $\delta\psi$ gives the stationary
Gross--Pitaevskii equation with the gravitational and three-particle interactions:

\begin{equation}
\left[-\frac{\hbar^2}{2m}\Delta_r-\mu+4\pi Gm^2V(r)\right]\psi+U_2\psi^3+U_3\psi^5=0,\quad
\psi(R)=0,
\end{equation}
where $R$ is defined as the (first) zero of $\psi(r)$ and determines a core boundary of BEC.

The further description of the system is conveniently carried out in terms of
dimensionless wave function $\chi$ and coordinate $\xi$: 

\begin{equation}
\psi(r)=\sqrt{n_0}\,\chi(\xi),\quad
r=r_0\,\xi,\quad
R=r_0\,\xi_B,
\end{equation}
where  $n_0$ and $r_0$ set the scale of the particle density and radius.

A normalization of the dimensionless wave function $\chi(\xi)$ defines the mean
particle density $n$ in the total volume $4\pi R^3/3$:

\begin{equation}\label{norm}
\sigma\equiv\frac{3}{\xi^3_B}\int_0^{\xi_B}\chi^2(\xi)\,\xi^2\,\rmd\xi=\frac{n}{n_0}.
\end{equation}
Empirically, the value of $\sigma$ is less than unity \cite{GKN20,GN21}.

The wave function $\chi(\xi)$ determines the local particle density
$\eta(\xi)=\chi^2(\xi)$ and is found from the field equation

\begin{equation}\label{FEq1}
\frac{1}{2}(\Delta_\xi+\nu)\chi-A\chi\Phi-Q\chi^3-B\chi^5=0,
\end{equation}
where dimensionless parameters of the model are

\begin{equation}\label{ours}
Q=4\pi r^2_0 a n_0,\quad
A=Q\,\frac{Gm^3 r^2_0}{\hbar^2 a},\quad
B=g\,Q\,\lambda^3n_0,\quad
\varepsilon_0=\frac{\hbar^2}{m r^2_0}\,n_0.
\end{equation}

The effective chemical potential $\nu$ plays the role of free parameter 
absorbing the gravitational potential in the center (at $\xi=0$):

\begin{equation}
\upsilon=-A\int_0^{\xi_B}\eta(\xi)\,\xi\,\rmd\xi,
\end{equation}
and it leads to the following expression for $\Phi(\xi)$:

\begin{equation}\label{FEq2}
\Phi(\xi)=-\frac{1}{\xi}\int_0^\xi \eta(s)\,s^2\,\rmd s+\int_0^{\xi}\eta(s)\,s\,\rmd s,\qquad
\Delta_\xi\Phi(\xi)=\eta(\xi).
\end{equation}

As usual \cite{LL9}, we associate $r_0$ with the {\em healing length}:

\begin{equation}\label{rcor}
r_0=\frac{1}{\sqrt{4\pi an_0}},
\end{equation}
what leads immediately to $Q=1$ in (\ref{ours}).
Under the diluteness conditions (\ref{naaa}) which guarantee applicability
of the given formalism, the following inequality holds:

\begin{equation}
r^3_0n_0\gg 1.
\end{equation}

Let us relate the characteristics of dark particles with the parameters of our model, using
the typical mass density value $\rho_0=10^{-20}\ \text{kg}/\text{m}^3$, that gives

\begin{equation}
n_0=\frac{\rho_0}{m}.
\end{equation}

Then, substituting it and (\ref{rcor}) into (\ref{ours}), we obtain the relations

\begin{eqnarray}
&&m=\left(\frac{g}{B}\right)^{1/4}m_0,\quad
m_0\equiv\left(\frac{\hbar^3}{c^3}\,\rho_0\right)^{1/4}=0.081\ \text{eV}/c^2,\\
&&a=\left(\frac{g}{4\pi AB}\right)^{1/2}\,\ell_{\text{P}},\quad
\ell_{\text{P}}\equiv\sqrt{\frac{G\hbar}{c^3}}=1.62\times10^{-35}\ \text{m},\\
&&r_0=\left(\frac{B}{g}\right)^{1/8}\,\left(\frac{A}{4\pi}\right)^{1/4}\,L,\quad
L\equiv\sqrt{\frac{m_0}{\rho_0\,\ell_{\text{P}}}}=9.44\times10^8\ \text{m}.
\end{eqnarray}

It is required that $mc^2\leq1\ \text{eV}$ accordingly to \cite{FMT08}.
For definiteness, let us fix $g=4.73\times10^{-83}$ and

\begin{equation}\label{opar}
Q=1,\quad
A=\pi^2\simeq 9.87,\quad
B=20,
\end{equation}
to obtain that $a=2.23\times10^{-78}\ \text{m}$, $m=10^{-22}\ \text{eV}/c^2$,
$r_0=0.818\ \text{kpc}$.

Of course, we admit a more suitable parametrization for concrete situations.
Besides, the core size
$R=r_0\xi_B$ and the mass density $\rho=\rho_0\sigma$ should be found by solving
the field equations for given restrictions. However, here we focus on
analyzing the physical effects without applying the model to specific objects. 

Equations (\ref{FEq1}), (\ref{FEq2}) are numerically integrated under the following
initial conditions: $\chi^\prime(0)=0$ and $\chi^{\prime\prime}(0)<0$.
As it was argued in \cite{GN21}, the finite initial value $\chi(0)=z$ lies
in the range $z_1<z<z_2$, where

\begin{equation}
z_1=\left[\sqrt{\left(\frac{3Q}{10B}\right)^2+\frac{\nu}{10B}}-\frac{3Q}{10B}\right]^{1/2},\quad
z_2=\left[\sqrt{\left(\frac{Q}{2B}\right)^2+\frac{\nu}{2B}}-\frac{Q}{2B}\right]^{1/2},
\end{equation}
and it should be minimal positive solution of equation:

\begin{equation}\label{inc}
\frac{A}{2}z^2-\left(5Bz^4+3Qz^2-\frac{\nu}{2}\right)\left(\frac{\nu}{2}-Qz^2-Bz^4\right)=0.
\end{equation}
The absence of such a solution $z$ at given $(A,Q,B,\nu)$ means that $\chi(\xi)=0$ everywhere.
It happens at $\nu<\nu_{\text{min}}$ where $\nu_{\text{min}}(A,Q,B)$ is found numerically. 

Studying the statistical properties of matter, one uses the part of oscillating
wave function $\chi(\xi)$ in the range $\xi\in[0;\xi_B]$, limited by its
first zero $\chi(\xi_B)=0$. The value of $\xi_B$ determines the total radius $R=r_0\xi_B$
of the system and depends on the set of parameters $(A,Q,B,\nu)$.
Let us write down the internal energy density ${\cal E}$ and the local pressure ${\cal P}$
in dimensionless units:

\begin{equation}\label{ED}
{\cal E}(\xi)=\frac{Q}{2}\eta^2(\xi)+\frac{B}{3}\eta^3(\xi),\quad
{\cal P}(\xi)=\frac{Q}{2}\eta^2(\xi)+\frac{2}{3}B\eta^3(\xi).
\end{equation}

The statistical mean of any local characteristic is defined by the integral:

\begin{equation}
\langle(\ldots)\rangle=\frac{3}{\xi^3_B}\int_0^{\xi_B}(\ldots)\xi^2\rmd\xi.
\end{equation}

In general, there are two limiting regimes of the model (\ref{FEq1}), (\ref{FEq2}):
i) the Thomas--Fermi (TF) approximation, when the terms of quantum fluctuations like
$(\partial_\xi\chi)^2$ and $\Delta_\xi\chi$ are neglected; ii) the regime of the standing
spherical wave, when the chemical potential $\nu$ is large enough and/or the interaction
terms are negligibly small in (\ref{FEq1}).

Since our model generalizes the well-studied model with pair interaction ($B=0$)
in the Thomas--Fermi approximation \cite{Harko2011}, it is useful to indicate 
its basic equations and the values of the parameters. The set of this model
equations contains Equation~(\ref{FEq2}) and

\begin{equation}
\frac{\nu}{2}-A\Phi-Q\eta=0.
\label{HM}
\end{equation}

Solution to these equations is given (see \cite{Harko2011,Naz20}) by

\begin{equation}
\eta(\xi)=\eta_c\,j_0(\kappa\xi),\quad
\kappa=\sqrt{\frac{A}{Q}},\quad
\eta_c=\frac{\nu}{2Q},
\label{HM2}
\end{equation}
where $j_0(z)=\sin{z}/z$ is the spherical Bessel function.
In fact, the central density $\eta_c$ is regarded as a free parameter
of the model instead of $\nu$.

Limiting the system radius by the value $r=r_0$ (or $\xi_B=1$), one derives
that $\kappa=\pi$. Hence, for arbitrary $\nu>0$, the parameters of such
a model can be fixed as $Q=1$ and $A=\pi^2$. This confirms our choice of
the value of $A$ in (\ref{opar}). 

Note that close parameter values, $Q\sim1.36$ and $A=10$, were used
in \cite{GN21}. However, taking into account the quantum fluctuations 
$\Delta_\xi\chi$ and the three-particle interaction regulated by parameter
$B$, the system radius $\xi_B>1$ and the central density $\eta_c=\chi^2(0)$
reveal nonlinear dependence on $\nu$ (see Figure~\ref{XI} below and
Figure~3 in \cite{GKN20}).

Let us note the relation $r_0(a,m)$:

\begin{equation}
r_0=\pi\sqrt{\frac{\hbar^2a}{Gm^3}},
\end{equation}
which is discussed in detail in \cite{Harko2011} to outline physically reasonable
values of $a$ and $m$ by describing the dark matter of dwarf galaxies.

In contrast with the Thomas--Fermi approximation, equation (\ref{FEq1}) admits
the limiting regime of the standing wave, implied by the equation:

\begin{equation}
(\Delta_\xi+k^2)\chi(\xi)=0,\qquad
k^2=\nu-w>0,
\end{equation}
where $w$ is a correction due to small interaction, which is not derived here.

\begin{figure}[htbp]
\begin{center}
\includegraphics[width=8cm,angle=0]{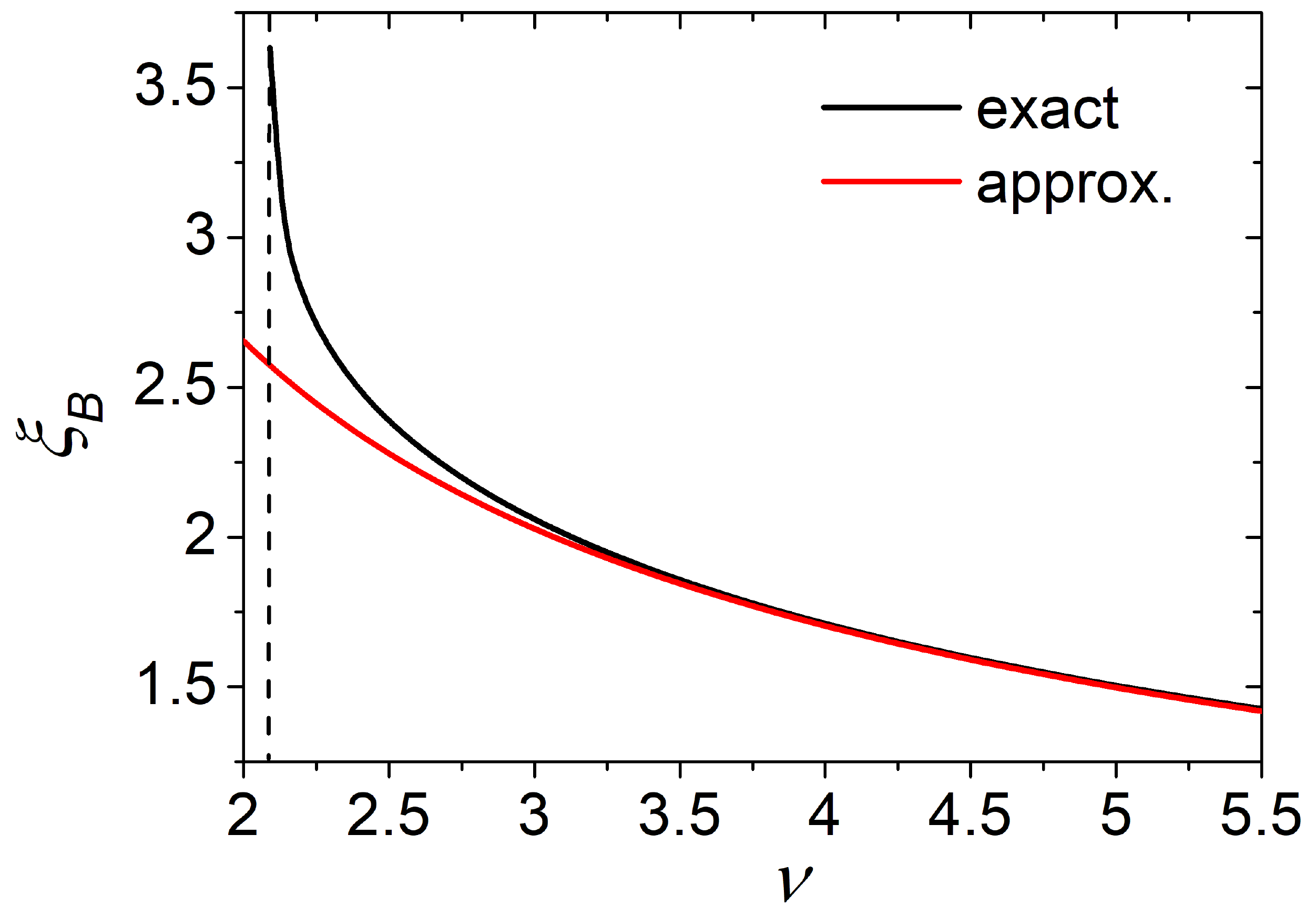}
\end{center}

\caption{\label{XI}\small Dependence of the dimensionless radius $\xi_B$
(the first zero of $\chi(\xi)$) on effective chemical potential $\nu$.
Here $A=\pi^2$, $Q=1$, $B=20$, $\nu_{\text{min}}\simeq2.09$.
The approximation (\ref{Xapp}) is built at $w=0.6$. The chemical
potential $\mu=0$ at $\nu\simeq2.263$, defining the lower limit of
 applicability of the approximation.}
\end{figure}

This equation has a simple analytic solution:

\begin{equation}\label{Xapp}
\chi(\xi)=\chi_0\,j_0\left(\pi\frac{\xi}{\xi_B}\right),\qquad
\xi_B=\frac{\pi}{\sqrt{\nu-w}}.
\end{equation}
Obvious merging of the approximate function $\xi_B(\nu)$ and
the numerically found one at large $\nu$ is shown in Figure~\ref{XI} for
the fixed values of $Q$ and $A$.

Besides, the form of wave function (\ref{Xapp}) guarantees that
$\langle\eta^n(\xi)\rangle=\eta^{2}_c\,c_n$, where $\eta_c\equiv\chi^2_0$,
and the numeric coefficients $c_n$ are independent of $\nu\gg\nu_{\text{min}}$.
Although the exact situation is only qualitatively the same, as we shall
see in Figure~\ref{Cs} below, this effect excludes a possibility of the
composites formation because of destruction of many-particle states at large
both $\nu$ and particle momentum $k=\sqrt{\langle(\partial_\xi\chi)^2\rangle/\sigma}$.
As it is easily seen, the averages of $\eta^n$ with similar properties
are obtained for the solution (\ref{HM2}) in the Thomas--Fermi approximation.

Referring to \cite{GN21}, we especially note that the successful 
attempts of reproducing rotational curves of dwarf galaxies require
the use of $\nu\simeq\nu_{\text{min}}$, what can be associated with
a ground state of the model (\ref{FEq1}), (\ref{FEq2}).

Moreover, the two regimes of ``small'' and ''large'' $\nu$ are
naturally separated by condition $\mu=0$ for the physical chemical
potential or its dimensionless value $u\equiv \nu/2-A\upsilon=0$.
In other words, the regime of large $\nu$ corresponds to $u>0$,
while the bound states are described at $u<0$. As we have seen
in \cite{GN21,GKN20}, the point $u=0$ determines parameters of
the first-order phase transition in the system with three-particle
interaction.

After matching the parameters of our model and the reference one, we would also like
to compare some dependencies. We start by examining the mean particle density $\sigma$
as a function of the central density $\eta_c$, at large values of which
the transformation of particles can occur.

As already shown in \cite{Harko2011} for model (\ref{HM})-(\ref{HM2}) with
$\xi_B=1$, the relation between $\sigma$ and $\eta_c$ is given by

\begin{equation}\label{HM3}
\sigma=\frac{3}{\pi^2}\,\eta_c.
\end{equation}

\begin{figure}[htbp]
\begin{center}
\includegraphics[width=13.3cm,angle=0]{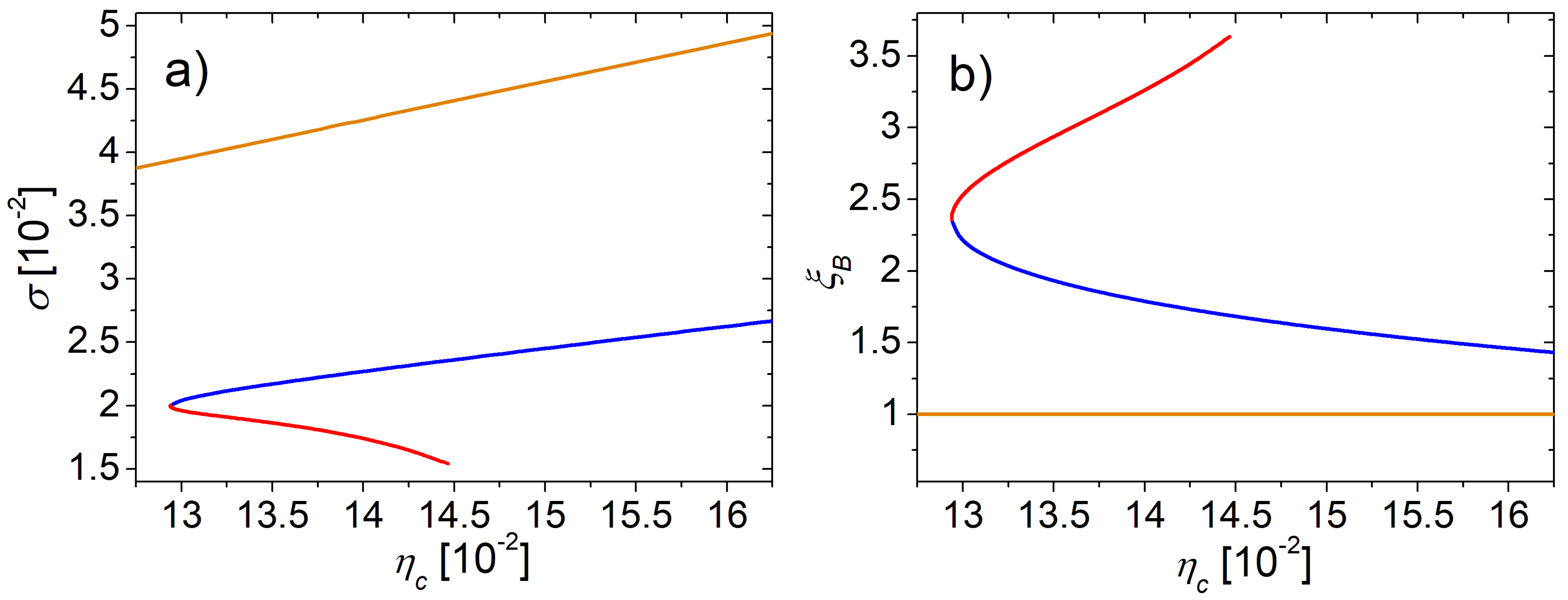}
\end{center}
\caption{\label{dens}\small Mean particle density $\sigma$ (\textbf{a}) and system radius $\xi_B$ (\textbf{b})
as functions of central density $\eta_c$ in two models. In both panels, the orange line corresponds
to the model with pair interaction in the Thomas--Fermi approximation. The red and blue branches of
the single curve correspond to the model that takes into account both quantum fluctuations and
three-particle interaction.}
\end{figure}

A completely different relation is expected for our model (\ref{FEq1}).
The dependencies of interest are shown in Figure~\ref{dens}a.

The first difference between the model with pair interaction in the Thomas--Fermi
approximation (\ref{HM})-(\ref{HM2}) and our model which takes into account both
quantum fluctuations and three-particle interaction is the presence of the minimal
admissible central density in the latter, which for the given model parameters
is equal to $\eta^*_c\simeq0.1294$. 

At $\eta_c\geq\eta^*_c$, there are two branches of the function $\sigma(\eta_c)$
for the model (\ref{FEq1}), which are colored in blue and red in
Figure~\ref{dens}a. The behavior of the blue branch is similar to the orange line
described by (\ref{HM3}), while the red curve looks abnormal due to three-particle
repulsion. To explain the discrepancy between the mean density curves colored in
blue and orange as well as the red branch, let us turn to the dependence of
the size $\xi_B$ on $\eta_c$. 

We see in Figure~\ref{dens}b that an increase in the system size $\xi_B$ along red branch
is accompanied by a decrease in the corresponding mean density in Figure~\ref{dens}a.
That is, the red sectors in Figures \ref{dens}a and \ref{dens}b describe quite dilute matter.

Since the blue curve in Figure~\ref{dens}b describes a system with a larger radius $\xi_B$
than the orange line for the reference model, this is consistent with the lower density in
blue compared to the mean density in orange in Figure~\ref{dens}a. As $\eta_c$ grows,
the blue curve $\xi_B(\eta_c)$ tends to $\xi_B=1$. Therefore, the blue curves do show
a denser state of matter.

On the other hand, the existence of composites in this dense matter becomes problematic
due to the mutual influence leading to their dissociation into particles.
Assuming that the interactions still guarantee the formation of composites for
a short time, which is similar to oscillations, we will consider such a process
in the next Section using the model of quantum random walk. 

To test our assumptions, we introduce the ratios of the dimensionless means:

\begin{equation}
C_1=\frac{\langle\eta^2\rangle}{\langle\eta\rangle^2},\quad
C_2=\frac{\langle\eta^3\rangle}{\langle\eta\rangle^3},\quad
C_3=\frac{\langle\eta^3\rangle}{\langle\eta\rangle^2\langle\eta\rangle},
\label{corr}
\end{equation}
where the particle density $\eta(\xi)=\chi^2(\xi)$ is formed from the solution $\chi(\xi)$
of (\ref{FEq1}).

\begin{figure}[htbp]
\begin{center}
\includegraphics[width=8.8cm,angle=0]{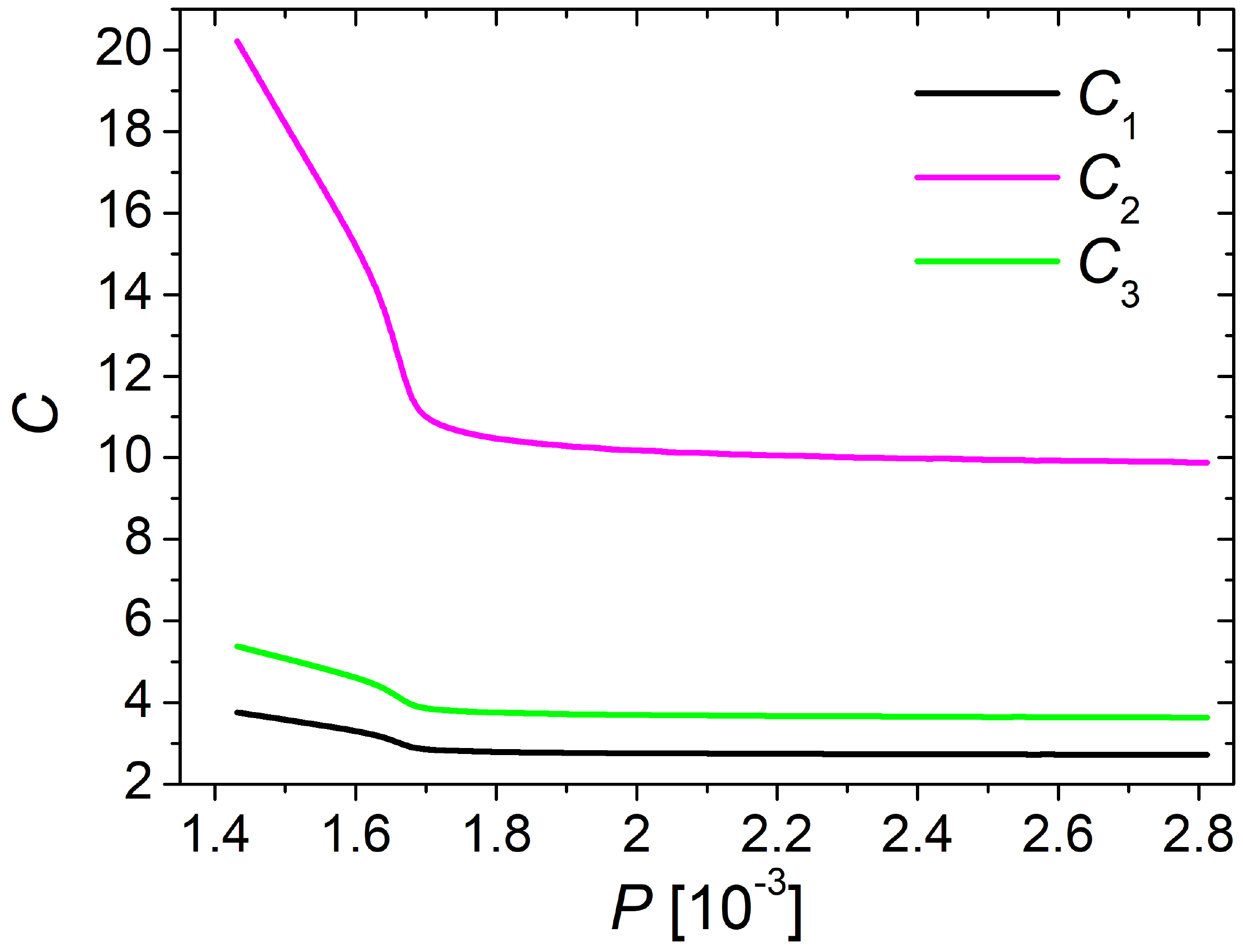}
\end{center}
\caption{\label{Cs}\small Behavior of the ratios $C_{1,2,3}$ of mean values
$\langle\eta^3\rangle$, $\langle\eta^2\rangle$ and $\langle\eta\rangle$
[see (\ref{corr})] with varying internal pressure $P$.}
\end{figure}

To examine the behavior of these quantities, found with the same parameters $(A,Q,B)$
as before, but varying $\nu$, we use the mean internal pressure $P$ (see \cite{GN21}):

\begin{equation}\label{press}
P=\frac{Q}{2}\langle\eta^2\rangle+\frac{2}{3}B\langle\eta^3\rangle.
\end{equation}

The quotients $C_{1,2,3}(P)$, presented in Figure~\ref{Cs}, reveal the splitting and equivalence:

\begin{equation}
\langle\eta^2\rangle\sim\langle\eta\rangle^2, \quad
\langle\eta^3\rangle\sim\langle\eta\rangle^3, \quad
\langle\eta^3\rangle\sim\langle\eta^2\rangle\langle\eta\rangle,
\end{equation}
described by almost horizontal plateau at relatively high pressure $P>1.8\times10^{-3}$.
(To obtain the value of pressure and energy density in physical units, the dimensionless
quantity should be multiplied by $\varepsilon_0=\hbar^2 n_0/(mr^2_0)$, see (\ref{ours}).)
This means that two and three-particle configurations behave as two and three isolated
particles, respectively. The reasons of such a behavior are explained above by considering
the standing wave solution (\ref{Xapp}). The transition to this regime is, generally speaking,
a first-order phase transition described in \cite{GKN20,GN21}. 

\begin{figure}[htbp]
\begin{center}
\includegraphics[width=13.2cm,angle=0]{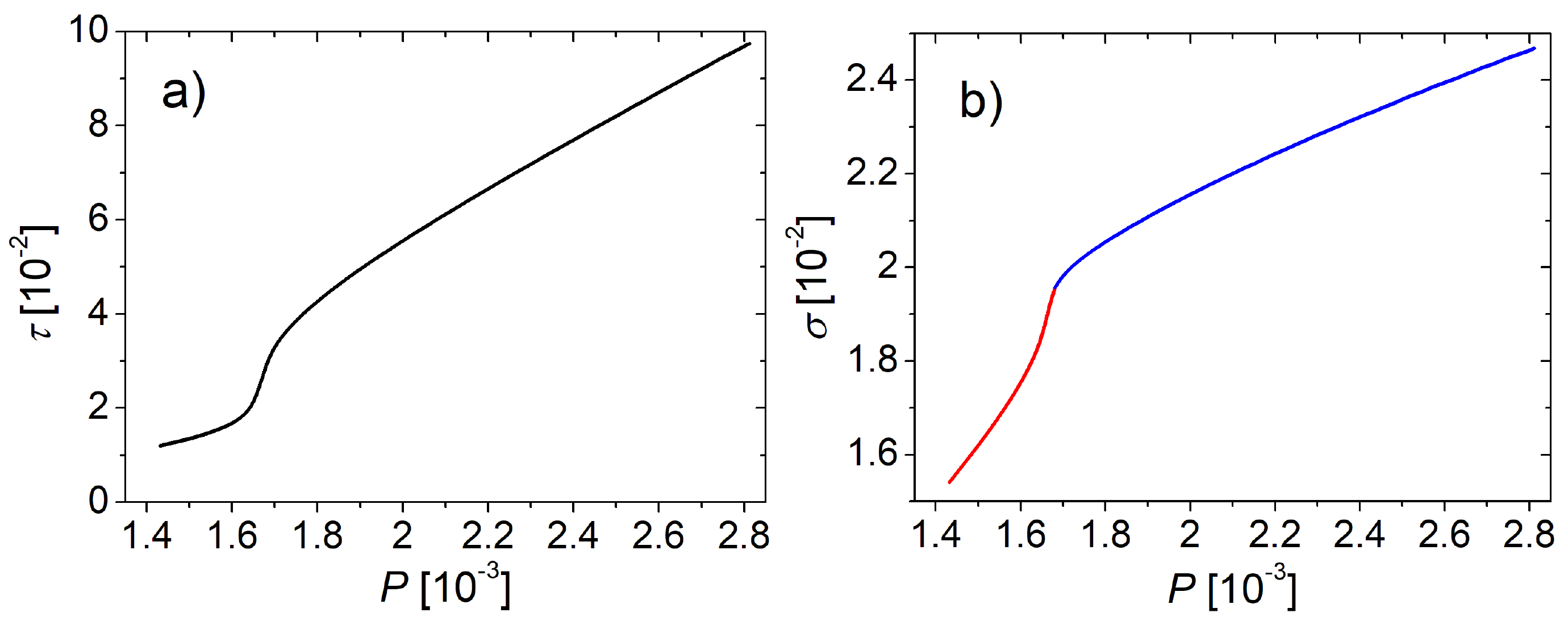}
\end{center}
\caption{\label{tau}\small Dependence of the measure of quantum fluctuations $\tau$ (\textbf{a})
and mean density (\textbf{b}) on internal pressure $P$.}
\end{figure}

It is natural to associate the instability of many-particle configurations (composites)
with quantum fluctuations in the system. Indeed, the quantity
$\tau=\langle(\partial_\xi\chi)^2\rangle$
undergoes a jump in Figure~\ref{tau}a when passing to the pressure interval of interest.
Moreover, at $P>1.8\times10^{-3}$, the almost linear proportionality between $P$ and $\tau$
resembles the relation between pressure and temperature of an ideal gas.

In summary, we can conclude that the high pressure regime does not provide the formation
of composites in BEC, although the particle density increases with the pressure,
as shown in Figure~\ref{tau}b. Besides, the inequality $nr^3_0\gg1$ is fulfilled in BEC
due to large healing length $r_0$, which determines the BEC core radius~\cite{LL9}.
Despite the strong correlations, characterized by $r_0$, the diluteness condition
$na^3<1$ persists. 

\section{Quantum Random Walk Based Treatment}

Based on the predictions of the previous Section, we would like to consider
the instability of two- and three-particle states in dense matter within the framework
of a simple probabilistic model. 
In the case of an elementary system of three particles (monomers, denoted M), which are confined
within the radius of interactions leading to the formation of a dimer (D) and a trimer (T),
there is a possibility of oscillations, namely, a continuous transition between three states:

\begin{eqnarray}
&\psi_1=|3\rangle_{\mathrm{M}}\bigotimes|0\rangle_{\mathrm{D}}\bigotimes|0\rangle_{\mathrm{T}},\qquad
\psi_2=|1\rangle_{\mathrm{M}}\bigotimes|1\rangle_{\mathrm{D}}\bigotimes|0\rangle_{\mathrm{T}},&\\
&\psi_3=|0\rangle_{\mathrm{M}}\bigotimes|0\rangle_{\mathrm{D}}\bigotimes|1\rangle_{\mathrm{T}},&\nonumber
\end{eqnarray}
representing the unbound state and two bound ones. In the absence of external factors,
the reversibility of microscopic processes is supported by the equality of the probabilities of
forming and decaying of the composites. Of course, the known lifetimes of each (metastable) state
could point out the dominant configuration of the particles.
However, having permanently interacting particles, we will neglect the probability of their adhesion.
Nevertheless, it seems possible to estimate the probabilities of states over the period of oscillations. 

Thus, we consider a toy model with three identical monomers, which is described by
the wave function $\Psi(t)$:

\begin{equation}
\Psi(t)=\sum\limits_{i=1}^3a_i(t)\psi_i,\qquad \sum\limits_{i=1}^3|a_i(t)|^2=1,
\end{equation}
that evolves in time $t$ according to the Schr\"odinger equation
with initial state (at $t=0$) assumed to be $\Psi_0=\psi_1$.

Then we define the fractions $\Omega_i(t)=|a_i(t)|^2$ and analyze
their means $\overline{\Omega}_i$ over a period of time. 

Evolution of the initial state,

\begin{equation}\label{incon}
|\Psi_0\rangle=
\begin{pmatrix}
1\\
0\\
0
\end{pmatrix}\xrightarrow{\quad t \quad}
|\Psi(t)\rangle=
\begin{pmatrix}
a_1(t)\\
a_2(t)\\
a_3(t)
\end{pmatrix},
\end{equation}
is given by the Schr\"odinger equation with Hamiltonian $\hat H=\hat H_0+\hat U$,
where $\hat H_0=\mathrm{diag}(\varepsilon_1,\varepsilon_2,\varepsilon_3)$:

\begin{equation}\label{Sch1}
\rmi\hbar\frac{\rmd}{\rmd t}|\Psi\rangle=\hat H|\Psi\rangle.
\end{equation}

Actually, the matrix $\hat H$ with constant elements determines quantum random walk
(a Markov process). Analytically, solution of (\ref{Sch1}) for the wave function
can be immediately represented (and found) in the form:

\begin{equation}\label{Smat}
|\Psi(t)\rangle=\hat S(t)|\Psi_0\rangle,\qquad
\hat S(t)=\exp{\left(-\frac{\rmi}{\hbar}\hat Ht\right)}.
\end{equation}

Neglecting spatial evolution of localized particles and time delay in each state,
we omit $\hat H_0$ and remain with the matrix of admissible transitions: 

\begin{equation}
\hat U=
\begin{pmatrix}
0&U_{-1}&U_{-3}\\
U_1&0&U_{-2}\\
U_3&U_2&0
\end{pmatrix},
\end{equation}
where the quantities $U_{\pm i}$ play a role of transition amplitudes,
and we put $U_{-i}=U_i\in\mathbb{R}$ to make $\hat U$ self-adjoint with
real elements.

The conserved energy of the system is

\begin{equation}
E\equiv\langle\Psi(t)|{\hat H}|\Psi(t)\rangle
=U_1a^*_2(t)a_1(t)+U_2a^*_3(t)a_2(t)+U_3a^*_1(t)a_3(t)+\text{c.c.}
\end{equation}

Introducing auxiliary quantities:

\begin{equation}
U^{(3)}=U_1U_2U_3,\qquad U^{(2)}=\frac{1}{3}(U^2_1+U^2_2+U^2_3),\\
\end{equation}
three eigenvalues of $\hat U$, which are roots of equation

\begin{equation}
0=\det\left(\hat U-\lambda\hat I\right)=-\lambda^3+3\lambda U^{(2)}+2U^{(3)},
\end{equation}
are found to be

\begin{eqnarray}
&\lambda_1=A+B,\qquad \lambda_2=\rme^{-2\pi\rmi/3}A+\rme^{2\pi\rmi/3}B,&\nonumber\\
&\lambda_3=\rme^{2\pi\rmi/3}A+\rme^{-2\pi\rmi/3}B,&
\end{eqnarray}
where

\begin{equation}\label{AB}
A=\left[U^{(3)}+\sqrt{\left(U^{(3)}\right)^2-\left(U^{(2)}\right)^3}\right]^{1/3},\ 
B=\frac{U^{(2)}}{A}.
\end{equation}

General solution to equation (\ref{Sch1}) is presented in the form:

\begin{equation}\label{gensol}
|\Psi(t)\rangle=\sum\limits_{i=1}^3C_i\,{\bf u}_i\,\exp{\left(-\rmi\frac{\lambda_it}{\hbar}\right)},\qquad
{\bf u}_i=
\begin{pmatrix}
\lambda^2_i-U^2_2\\
\lambda_iU_1+U_2U_3\\
\lambda_iU_3+U_1U_2
\end{pmatrix},
\end{equation}
where ${\bf u}_i$ are the non-normalized eigenvectors of matrix $\hat U$;
$C_i$ are the constants, determined by initial condition $|\Psi(0)\rangle=|\Psi_0\rangle$
from (\ref{incon}):

\begin{equation}
C_1=\rmi\frac{\sqrt{3}}{9}\frac{\lambda_2-\lambda_3}{A^3-B^3},\quad
C_2=\rmi\frac{\sqrt{3}}{18}\frac{\lambda_3-\lambda_2-3\lambda_1}{A^3-B^3},\quad
C_3=\rmi\frac{\sqrt{3}}{18}\frac{\lambda_3-\lambda_2+3\lambda_1}{A^3-B^3}.
\end{equation}

We omit the study of time dependence in three-parameter case (\ref{gensol})
and restrict ourselves to a particular case, setting $U_3=0$ in (\ref{AB}), when 

$$
A=3^{-1/2}\,\rme^{\rmi\pi/6}\sqrt{U^2_1+U^2_2}.
$$

We easily obtain the spectrum:

\begin{equation}\label{spec2}
\lambda_1=\sqrt{U^2_1+U^2_2},\ \ \lambda_2=0,\ \ \lambda_3=-\sqrt{U^2_1+U^2_2}.
\end{equation}

Physically, these quantities determine the energy levels in the system and
indicate the presence of the unbound state of 3M (three monomers) with $\lambda_1$,
the bound one of T with $\lambda_3$, and a transient one with $\lambda_2$,
associated with M+D.

It is useful to parametrize $U_1$ and $U_2$ by $s$ and $T$ as

\begin{equation}\label{Uparam}
U_1=J\cos{s},\quad U_2=J\sin{s},\quad J=\hbar\omega=\frac{2\pi\hbar}{T}.
\end{equation}

Parameter $T$ defines the time scale of the problem, while the quantities $\cos{s}/T$
and $\sin{s}/T$ represent the rates of the processes.

Under the imposed conditions, the components of the matrix $\hat S(t)$ are easily calculated
and yield the instant fractions:

\begin{eqnarray}
&\Omega_{\mathrm{M}}(t,s)=\left[\cos^2{s}\,\cos{(\omega t)}+\sin^2{s}\right]^2,\qquad
\Omega_{\mathrm{D}}(t,s)=\left[\cos{s}\,\sin{(\omega t)}\right]^2,&\\
&\Omega_{\mathrm{T}}(t,s)=\left[\cos{s}\,\sin{s}\,\left(\cos{(\omega t)}-1\right)\right]^2.&\nonumber
\end{eqnarray}

To obtain this, we can also use in (\ref{gensol}) that

\begin{equation}
C_1{\bf u}_1=\frac{1}{2}
\begin{pmatrix}
\cos^2{s}\\
\cos{s}\\
\cos{s}\,\sin{s}
\end{pmatrix},\ \
C_2{\bf u}_2=
\begin{pmatrix}
\sin^2{s}\\
0\\
-\cos{s}\,\sin{s}
\end{pmatrix},\ \
C_3{\bf u}_3=\frac{1}{2}
\begin{pmatrix}
\cos^2{s}\\
-\cos{s}\\
\cos{s}\,\sin{s}
\end{pmatrix}.
\end{equation}

Note that configuration with $\Omega_{\mathrm{M}}=0$ and $\Omega_{\mathrm{T}}>\Omega_{\mathrm{D}}$
is realized at the time moment (during the first period)

\begin{equation}
t_1=\frac{T}{2}-\frac{T}{2\pi}\arccos{\left(\tan^2{s}\right)},
\end{equation}
when $s$ tends to $\pi/4$ from below.

\begin{figure}[htbp]
\begin{center}
\includegraphics[width=9cm,angle=0]{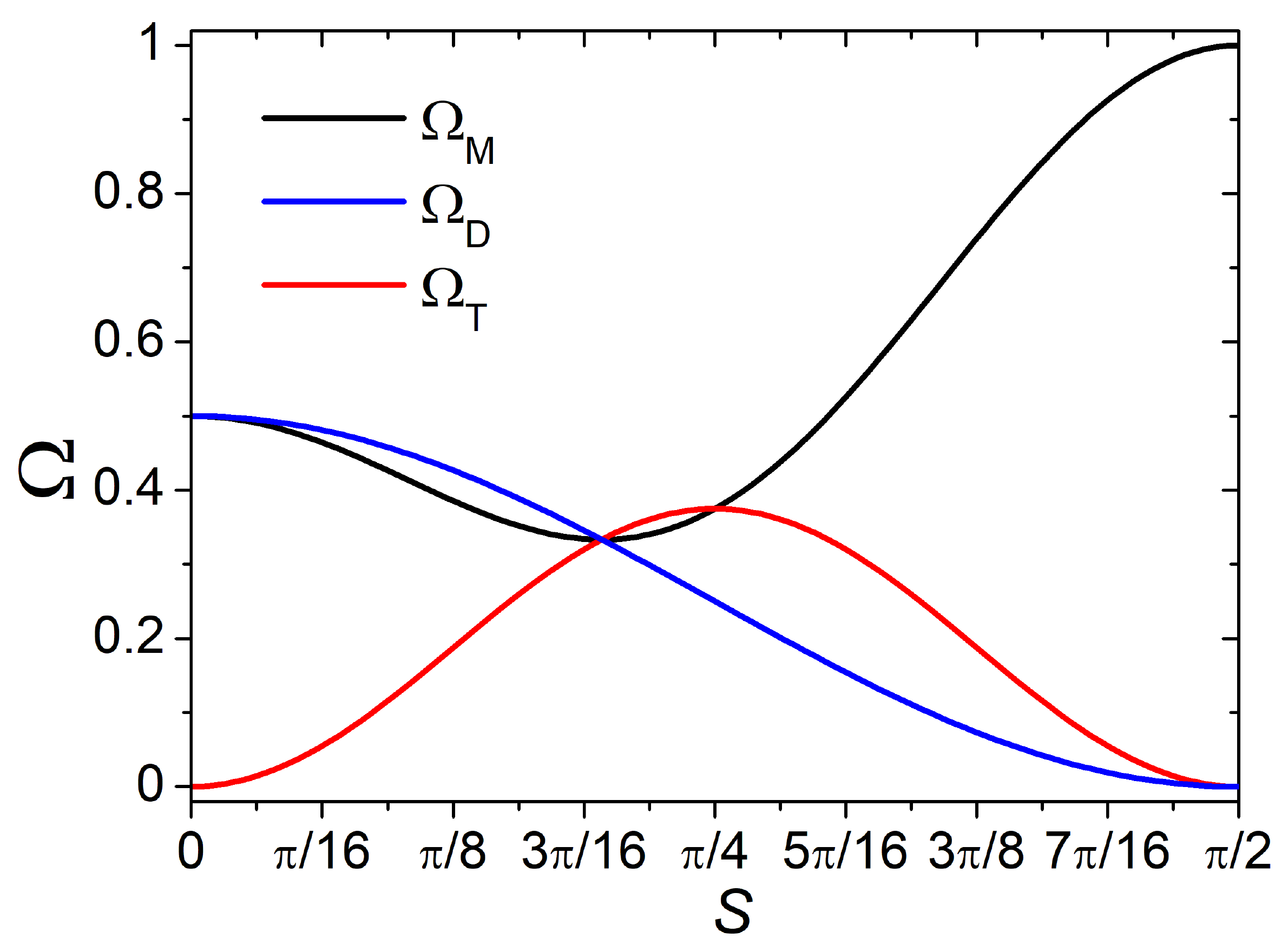}
\end{center}
\caption{\label{QRW}\small Time-averaged fractions $\overline{\Omega}_\alpha$ as functions of
parameter $s=\mathrm{arctan}\left(U_2/U_1\right)$ at $U_3=0$.}
\end{figure}

Averaging $\Omega_\alpha(t)$ over the period $T$, we arrive at

\begin{eqnarray}
&&\overline{\Omega}_{\mathrm{M}}(s)=\frac{3}{2}\cos^4{s}-2\cos^2{s}+1,\nonumber\\
&&\overline{\Omega}_{\mathrm{D}}(s)=\frac{1}{2}\cos^2{s},\quad
\overline{\Omega}_{\mathrm{T}}(s)=\frac{3}{2}\cos^2{s}\,\sin^2{s}.
\end{eqnarray}

Comparison of these fractions for different $s$ is made in Figure~\ref{QRW}. We see
a slight dominance of $\overline{\Omega}_{\mathrm{T}}$ in the range $s_c<s<\pi/4$,
while $\overline{\Omega}_{\mathrm{D}}$ prevails for $s<s_c$, where 
$s_c=\mathrm{arcsin}\left(\sqrt{3}/3\right)\simeq3.13\pi/16$, and
$\overline{\Omega}_{\mathrm{M}}(s_c)=\overline{\Omega}_{\mathrm{D}}(s_c)=\overline{\Omega}_{\mathrm{T}}(s_c)=1/3$.

Additionally, let us indicate the changes in the energy spectrum when $\hat H_0\not=0$.
Giving $\hat H_0=\mathrm{diag}(\mu,\mu,\mu)$, we get a shift in (\ref{spec2}):
$\lambda_i\to\lambda_i+\mu$. Including $\hat H_0=\mathrm{diag}(0,-2J\sinh(a),0)$ leads
to rescaling: $\lambda_1=J\rme^{-a}$, $\lambda_2=0$,  $\lambda_3=-J\rme^{a}$
[see (\ref{Uparam})]. Dynamics in these cases will be somewhat more complicated and
we omit it.

In summary, we note that in a three-particle system with permanent oscillations
between quantum states there exists the possibility of the composites dominance
in a long-time picture. This effect is due to the tuning of interaction
parameters. In general, the probability of detecting monomers (of dark matter)
increases at higher density and pressure in the model with few-particle interactions.

\section{Chemical Kinetics Consideration}

For the sake of simplicity, let us assume that the elementary constituents of
dark matter are identical spinless particles (monomers, or M). Due to an
interaction between them, the nature of which is not specified herein, we
assume the ability of formation of dimers~(D) and trimers~(T) in the ground
state. Therefore, we restrict ourselves to processes involving no more than three
monomers, taking into account also those contained in dimers and trimers. 
Attempting to describe the formation/decay of dimers and trimers, as well as their
fractions, we clarify some physical conditions. 

Account of the gravitational interaction, resulting in an inhomogeneous
distribution of particles, implies that the statistical characteristics of
the matter should, in principle, depend on space. We assume that the particle
density is highest in a core and tends to zero at halo periphery. However,
in regions of space that are small in astronomical scale, but large enough
to apply the statistical approach, the gravitational potential (together with
the chemical one) can be considered as constant one. This allows us to explore
the properties of quasi-homogeneous subsystems. However, simplifying
the problem by introducing regions of homogeneity with different particle densities,
the dominance of certain processes does depend on the ratio of the interparticle
distance to the radius of interaction (or scattering lengths).

The particle density in the core can be estimated as

\begin{equation}\label{n0}
n_0\simeq 5.61\cdot10^{15}\, \mathrm{m}^{-3}\,
\left[\frac{mc^2}{1\ \mathrm{eV}}\right]^{-1}
\left[\frac{\rho_0}{10^{-20}\ \mathrm{kg}\,\mathrm{m}^{-3}}\right],
\end{equation}
which is determined by the individual bosonic monomer mass $m\lesssim1$~eV and the typical
mass density $\rho_0$.

Let us consider possible reactions occurring simultaneously in each subsystem:

\begin{equation}\label{acts}
2\mathrm{M}\xrightleftharpoons[k_{-1}]{k_1}\mathrm{D},\qquad
\mathrm{M}+\mathrm{D}\xrightleftharpoons[k_{-2}]{k_2}\mathrm{T},\qquad
3\mathrm{M}\xrightleftharpoons[k_{-3}]{k_3}\mathrm{T},
\end{equation}
where the positive constants $k_i$ and $k_{-i}$ ($i=1,2,3$) are
related to the rates of the corresponding reactions and have the dimension
of inverse time.

The time evolution of concentrations $C_\alpha$ (the numbers of
particles of sort $\alpha=\mathrm{M},\mathrm{D},\mathrm{T}$ in a fixed volume)
can be described accordingly to the rules of chemical kinetics~\cite{Zam}:

\begin{eqnarray}
\frac{\rmd C_{\mathrm{M}}}{\rmd t}&=&-2k_1V_{\mathrm{2M}}C^2_{\mathrm{M}}
+2k_{-1}C_{\mathrm{D}}-k_2V_{\mathrm{MD}}C_{\mathrm{M}}C_{\mathrm{D}}\nonumber\\
&&+k_{-2}C_{\mathrm{T}}-3k_3V^2_{\mathrm{3M}}C^3_{\mathrm{M}}+3k_{-3}C_{\mathrm{T}},\nonumber\\
\frac{\rmd C_{\mathrm{D}}}{\rmd t}&=&k_1V_{\mathrm{2M}}C^2_{\mathrm{M}}-k_{-1}C_{\mathrm{D}}
-k_2V_{\mathrm{MD}}C_{\mathrm{M}}C_{\mathrm{D}}+k_{-2}C_{\mathrm{T}},\label{Kin}\\
\frac{\rmd C_{\mathrm{T}}}{\rmd t}&=&k_2V_{\mathrm{MD}}C_{\mathrm{M}}C_{\mathrm{D}}-k_{-2}C_{\mathrm{T}}
+k_3V^2_{\mathrm{3M}}C^3_{\mathrm{M}}-k_{-3}C_{\mathrm{T}},\nonumber
\end{eqnarray}
with the initial conditions at $t=0$: $C^{(0)}_{\mathrm{M}}>0$, $C^{(0)}_{\mathrm{D,T}}=0$;
for $V_{\mathrm{2M}}$, $V_{\mathrm{3M}}$, and $V_{\mathrm{MD}}$, see next proposition.

The above set of equations for $C_\alpha$ conserves in time the total number of monomers
in the subsystem of a fixed volume,

\begin{equation}
C_{\mathrm{M}}(t)+2C_{\mathrm{D}}(t)+3C_{\mathrm{T}}(t)=\mathrm{const},
\end{equation}
and implies that the elementary act depends on constants $k_{\pm i}$ and
interaction volumes $V_s\simeq a^3_s$, where $a_{\mathrm{2M}}$, $a_{\mathrm{3M}}$
and $a_{\mathrm{MD}}$ are the scattering lengths (for $s$-wave channel, in instance).
It is natural to require conditions for the successful performing of three forward
reactions:

\begin{eqnarray}
&&V_{\mathrm{2M}}C_{\mathrm{M}}\geq1, \quad V_{\mathrm{3M}}C_{\mathrm{M}}\geq1,\nonumber\\
&&V_{\mathrm{MD}}C_{\mathrm{M}}\geq1 \quad \mathrm{and}\quad V_{\mathrm{MD}}C_{\mathrm{D}}\geq1.
\label{na3}
\end{eqnarray}
They indicate the presence of at least one particle in each interaction volume $V_s$.
Failure to meet one of these conditions may lead to the exclusion of the corresponding
process from consideration. 

Associating $C_\alpha$ with the local particle density far from the core (of galactic halo)
center, in order to satisfy the imposed conditions (\ref{na3}), we require that
$n_0a^3_s\gg1$ for the scattering lengths $a_s$ in the model.

Strictly speaking, the validity of condition $na^3_s>1$ depends on both $n$ and $a_s$.
However, as shown above, an increase in monomer density $n$ (due to squeezing, for example)
correlates with the destruction of composites. Therefore, there
remains the only possibility of considering interactions with longer scattering lengths
$a_s$. Next, we evaluate the contribution of dimers and trimers to dark matter. 

To simplify further notations, we will also use

\begin{equation}
K_1=k_1V_{\mathrm{2M}},\ K_2=k_2V_{\mathrm{MD}},\
K_3=k_3V^2_{\mathrm{3M}},\ K_{-i}=k_{-i}.
\nonumber
\end{equation}

It is convenient to describe the subsystem evolution in terms of the fractions:

\begin{equation}
\Omega_\alpha=\frac{c_\alpha C_\alpha}{C_{\mathrm{M}}+c_{\mathrm{D}}C_{\mathrm{D}}+c_{\mathrm{T}}C_{\mathrm{T}}},\qquad
\Omega_{\mathrm{M}}+\Omega_{\mathrm{D}}+\Omega_{\mathrm{T}}=1,
\end{equation}
where $c_{\mathrm{M}}=1$, $c_{\mathrm{D}}=2$, $c_{\mathrm{T}}=3$.

Let us analyze the equilibrium fractions $\Omega^{\mathrm{eq}}_\alpha$ when
$\rmd C_\alpha/\rmd t=0$. For equilibrium concentrations (dropping the label `eq'),
we find $C_{\mathrm{D}}$ and $C_{\mathrm{T}}$ as functions of $C_{\mathrm{M}}$:

\begin{eqnarray}
&&C_{\mathrm{D}}=C^2_{\mathrm{M}}\,f_{\mathrm{D}}(C_{\mathrm{M}}),\qquad\quad
C_{\mathrm{T}}=C^3_{\mathrm{M}}\,f_{\mathrm{T}}(C_{\mathrm{M}}),\\
&&f_\alpha(x)=\frac{a^{(11)}_\alpha x+a^{(12)}_\alpha}{a^{(21)}_\alpha x+a^{(22)}_\alpha},
\quad A_\alpha=\left(\begin{array}{cc}
a^{(11)}_\alpha&a^{(12)}_\alpha\\
a^{(21)}_\alpha&a^{(22)}_\alpha
\end{array}\right),
\end{eqnarray}
where the elements of matrix $A_\alpha$ are formed from $K_{\pm i}$:

\begin{eqnarray}
&&A_{\mathrm{D}}=\left(\begin{array}{cc}
K_3K_{-2}&K_1(K_{-2}+K_{-3})\\
K_2K_{-3}&K_{-1}(K_{-2}+K_{-3})
\end{array}\right),\\
&&A_{\mathrm{T}}=\left(\begin{array}{cc}
K_2K_3&K_1K_2+K_3K_{-1}\\
K_2K_{-3}&K_{-1}(K_{-2}+K_{-3})
\end{array}\right).
\end{eqnarray}

For a subsystem in a fixed (unit) volume, the initial concentration $C^{(0)}_{\mathrm{M}}$
is related to the equilibrium one $C_{\mathrm{M}}$:

\begin{equation}\label{CC0}
C_{\mathrm{M}}+2C^2_{\mathrm{M}}\,f_{\mathrm{D}}(C_{\mathrm{M}})
+3C^3_{\mathrm{M}}\,f_{\mathrm{T}}(C_{\mathrm{M}})=C^{(0)}_{\mathrm{M}}.
\end{equation}
That means, the initial state $C^{(0)}_{\mathrm{M}}$ can be determined by using the $C_{\mathrm{M}}$.

Note the possibility to generalize our considerations by specifying the concentration
$C^{(q)}_\alpha$ of composites of sort $\alpha$ in certain quantum state $q$ at a given
temperature $T$, that is, the occupation of energy level $E^{(q)}_\alpha$ with the
degeneracy $g_q$. For the processes that do not violate the equilibrium distribution
of particles over states, one writes~\cite{Zam}:

\begin{equation}
C^{(q)}_\alpha=C_\alpha p_\alpha(q),\quad
p_\alpha(q)=\frac{g_q}{Z_\alpha}\exp{\left(-\frac{E^{(q)}_\alpha}{T}\right)},\quad
\sum\limits_q p_\alpha(q)=1,
\end{equation}
where $Z_\alpha$ is the partition function over states, and $C_\alpha$ is the total
concentration. From equation (\ref{Kin}), with account of Boltzmann factors $p_\alpha(q)$,
one can deduce the dependence of the reaction rates $k_{\pm i}$ on non-zero temperature.

At finite $K_{\pm i}$, there are two limiting regimes: i) $C_{\mathrm{M}}\to0$ (low concentration
of monomers) admits approximation: $C^{(0)}_{\mathrm{M}}=C_{\mathrm{M}}+O(C^2_{\mathrm{M}})$, and

\begin{eqnarray}
&&\Omega^{\mathrm{eq}}_{\mathrm{M}}=1-2\frac{K_1}{K_{-1}}\,C_{\mathrm{M}}+O(C^2_{\mathrm{M}}),\nonumber\\
&&\Omega^{\mathrm{eq}}_{\mathrm{D}}=2\frac{K_1}{K_{-1}}\,C_{\mathrm{M}}+O(C^2_{\mathrm{M}}),\label{i1}\\
&&\Omega^{\mathrm{eq}}_{\mathrm{T}}=3\frac{K_1K_2+K_3K_{-1}}{K_{-1}(K_{-2}+K_{-3})}\,C^2_{\mathrm{M}}+O(C^3_{\mathrm{M}});
\nonumber
\end{eqnarray}
ii) asymptotic expressions at large $C_{\mathrm{M}}$ yield:

\begin{eqnarray}
&&C^{(0)}_{\mathrm{M}}=\frac{3K_3}{K_{-3}}\,C^3_{\mathrm{M}}+o(C^2_{\mathrm{M}}),\nonumber\\
&&\Omega^{\mathrm{eq}}_{\mathrm{M}}=\frac{K_{-3}}{3K_3}\,C^{-2}_{\mathrm{M}}+O(C^{-3}_{\mathrm{M}}),\quad
\Omega^{\mathrm{eq}}_{\mathrm{D}}=\frac{2K_{-2}}{3K_2}\,C^{-1}_{\mathrm{M}}+O(C^{-2}_{\mathrm{M}}),\label{ii1}\\
&&\Omega^{\mathrm{eq}}_{\mathrm{T}}=1-\frac{2K_{-2}}{3K_2}\,C^{-1}_{\mathrm{M}}+O(C^{-2}_{\mathrm{M}}).
\nonumber
\end{eqnarray}

The formulas (\ref{i1})-(\ref{ii1}) are obtained by expanding up to the first non-trivial terms.
Thus, the normalization
$\Omega^{\mathrm{eq}}_{\mathrm{M}}+\Omega^{\mathrm{eq}}_{\mathrm{D}}+\Omega^{\mathrm{eq}}_{\mathrm{T}}=1$
is valid for $\Omega^{\mathrm{eq}}_\alpha$ in the same approximation.

As seen from (\ref{i1})-(\ref{ii1}), the case i) indicates a rather low probability of
obtaining composites D and T, while the mode ii) shows the dominant role of trimers (due to reaction
$\mathrm{M}+\mathrm{D}\rightarrow\mathrm{T}$) and the suppression of the monomer number
because of converting $3\mathrm{M}\rightarrow\mathrm{T}$. In this case we also see
that $C_{\mathrm{M}}\sim \left(C^{(0)}_{\mathrm{M}}\right)^{1/3}$.

Since the admissible value of $\Omega^{\mathrm{eq}}_{\mathrm{T}}$ varies in the interval $(0;1)$,
we can get $\Omega^{\mathrm{eq}}_{\mathrm{T}}$ higher than the expected threshold
$\Omega^{\mathrm{thr}}_{\mathrm{T}}\in(0;1)$ by requiring 
$C_{\mathrm{M}}>C^{\mathrm{thr}}_{\mathrm{M}}$. For a given $\Omega^{\mathrm{thr}}_{\mathrm{T}}$,
the value of $C^{\mathrm{thr}}_{\mathrm{M}}$ is determined by a real positive root $z$ of
the quartic equation: 

\begin{equation}\label{thr_eq}
3z^3f_{\mathrm{T}}(z)\left(1-\Omega^{\mathrm{thr}}_{\mathrm{T}}\right)
-\Omega^{\mathrm{thr}}_{\mathrm{T}}\left(z+2z^2f_{\mathrm{D}}(z)\right)=0.
\end{equation}

A similar problem arises when expressing $C_{\mathrm{M}}$ through $C^{(0)}_{\mathrm{M}}$.
Besides, one can also find the threshold value of $C^{(0)}_{\mathrm{M}}$
when $\Omega^{\mathrm{eq}}_{\mathrm{T}}=\Omega^{\mathrm{thr}}_{\mathrm{T}}$.  
To demonstrate some analytical solutions, we restrict ourselves to simpler special cases.

Let $K_{-1}=K_{-2}=K_3=0$. It means that the allowed processes are

\begin{equation}
3\mathrm{M}\xrightarrow{k_1}\mathrm{M}+\mathrm{D}\xrightarrow{k_2}\mathrm{T}
\xrightarrow{k_{-3}}3\mathrm{M}.
\end{equation}
The fractions in such a model is given by

\begin{eqnarray}
&&\Omega^{\mathrm{eq}}_{\mathrm{M}}=
\frac{K_2K_{-3}}{K_2K_{-3}+2K_1K_{-3}+3K_1K_2C_{\mathrm{M}}},
\nonumber\\
&&\Omega^{\mathrm{eq}}_{\mathrm{D}}=
\frac{2K_1K_{-3}}{K_2K_{-3}+2K_1K_{-3}+3K_1K_2C_{\mathrm{M}}},\\
&&\Omega^{\mathrm{eq}}_{\mathrm{T}}=
\frac{3C_{\mathrm{M}}K_1K_2}{K_2K_{-3}+2K_1K_{-3}+3K_1K_2C_{\mathrm{M}}},\quad
C_{\mathrm{M}}={\cal C}_+\left(C^{(0)}_{\mathrm{M}}\right),
\nonumber
\end{eqnarray}
where ${\cal C}_+$ is a positive solution among the two solutions ${\cal C}_\pm$ to
the quadratic equation obtained from (\ref{CC0}) under imposed restrictions:

\begin{equation}
{\cal C}_\pm=\sqrt{\frac{K_{-3}}{3K_1}}\left[\pm\sqrt{C^{(0)}_{\mathrm{M}}+{\cal C}}-\sqrt{{\cal C}}\right],
\quad
{\cal C}=\frac{K_{-3}}{12K_1}\left(1+2\frac{K_1}{K_2}\right)^2.
\end{equation}
Therefore, $C_{\mathrm{M}}\sim \sqrt{C^{(0)}_{\mathrm{M}}}$ within this model.

To guarantee $\Omega^{\mathrm{eq}}_{\mathrm{T}}\geq\Omega^{\mathrm{thr}}_{\mathrm{T}}$
for $\Omega^{\mathrm{thr}}_{\mathrm{T}}=0.5$, we arrive at the condition:

\begin{equation}\label{thr1}
C_{\mathrm{M}}\geq\frac{K_{-3}}{3K_1}+\frac{2K_{-3}}{3K_2}\quad\mathrm{or}\quad
C^{(0)}_{\mathrm{M}}=4{\cal C}\,\gamma(\gamma+1),\quad \gamma\geq1,
\end{equation}
with definite $\gamma$.

Rewriting this as

\begin{equation}
\frac{1}{3}\frac{k_{-3}}{k_1}\frac{1}{C_{\mathrm{M}}a^3_{\mathrm{2M}}}
+\frac{2}{3}\frac{k_{-3}}{k_2}\frac{1}{C_{\mathrm{M}}a^3_{\mathrm{MD}}}\leq1,
\end{equation}
one has two auxiliary inequalities: $C_{\mathrm{M}}a^3_{\mathrm{2M}}\geq k_{-3}/k_1$
and $C_{\mathrm{M}}a^3_{\mathrm{MD}}\geq k_{-3}/k_2$, which are obviously satisfied at
$k_{-3}/k_1\leq1$, $k_{-3}/k_2\leq1$, and $C_{\mathrm{M}}a^3_s\geq 1$ from (\ref{na3}).

Now let $K_3=K_{-3}=0$ in (\ref{acts}). Denoting $\varkappa_i=K_{-i}/K_i$, we obtain the fractions:

\begin{eqnarray}
&&\Omega^{\mathrm{eq}}_{\mathrm{M}}=
\frac{\varkappa_1\varkappa_2}{\varkappa_1\varkappa_2+2\varkappa_2C_{\mathrm{M}}+3C^2_{\mathrm{M}}},
\quad
\Omega^{\mathrm{eq}}_{\mathrm{D}}=
\frac{2\varkappa_2C_{\mathrm{M}}}{\varkappa_1\varkappa_2+2\varkappa_2C_{\mathrm{M}}+3C^2_{\mathrm{M}}},
\nonumber\\
&&\Omega^{\mathrm{eq}}_{\mathrm{T}}=
\frac{3C^2_{\mathrm{M}}}{\varkappa_1\varkappa_2+2\varkappa_2C_{\mathrm{M}}+3C^2_{\mathrm{M}}},
\quad
C_{\mathrm{M}}={\cal C}_0\left(C^{(0)}_{\mathrm{M}}\right),
\end{eqnarray}
where ${\cal C}_0$ denotes a real solution among three roots ${\cal C}_m$ ($m=0,\pm1$) of a cubic equation:

\begin{eqnarray}
&&\hspace*{-6mm}
{\cal C}_m={\cal A}\,\rme^{\rmi\frac{2\pi}{3}m}-\frac{{\cal A}^{(2)}}{{\cal A}}\,\rme^{-\rmi\frac{2\pi}{3}m}-\frac{2}{9}\varkappa_2,
\quad
{\cal A}=\left[{\cal A}^{(3)}+\sqrt{\left({\cal A}^{(3)}\right)^2+\left({\cal A}^{(2)}\right)^3}\right]^{1/3},\\
&&\hspace*{-6mm}
{\cal A}^{(2)}=\frac{\varkappa_2}{9}\left(\varkappa_1-\frac{4}{9}\varkappa_2\right),
\quad
{\cal A}^{(3)}=\frac{\varkappa_1\varkappa_2}{6}
\left[C^{(0)}_{\mathrm{M}}+\frac{2}{9}\varkappa_2\left(1-\frac{8}{27}\frac{\varkappa_2}{\varkappa_1}\right)\right].
\end{eqnarray}
To provide $\Omega^{\mathrm{eq}}_{\mathrm{T}}\geq0.5$, we demand 
$3C^2_{\mathrm{M}}\geq\varkappa_1\varkappa_2+2\varkappa_2C_{\mathrm{M}}$ [see (\ref{thr_eq})],
or, equivalently,

\begin{equation}\label{thr2}
C_{\mathrm{M}}\geq\frac{\varkappa_2}{3}\left(1+\sqrt{1+3\frac{\varkappa_1}{\varkappa_2}}\right).
\end{equation}
Therefore, $C_{\mathrm{M}}a^3_{\mathrm{MD}}>k_{-2}/k_2$, where $k_{-2}/k_2\leq1$ to permit the synthesis of T.
Thus, condition (\ref{thr2}) is not stronger than (\ref{na3}).

Summarizing, the applied analysis enables the leading role of composites under the derived conditions
(\ref{na3}) and the definite values of monomer density $C^{(0)}_{\mathrm{M}}$ and $C_{\mathrm{M}}$
[see (\ref{thr1}) and (\ref{thr2})]. However, the mechanism of the formation of composites
needs detailed study.

\section{Discussion}

In this paper, we have considered three models that deal with the problem of formation and
stability of the two- and three-particle composites in dark matter.

Starting with the Bose-condensate dark matter model with respective self-interaction
on the base of the Gross--Pitaevskii equation extended by an additional three-particle
interaction, we find that the few-particle 
complexes can be formed at relatively low density and pressure, when quantum
fluctuations and the transfer momentum are small enough and cannot destroy the condensate
and the desired structures. Such a conclusion is deduced on the base of calculating
the averages of the powers of the local density $\eta$ and thermodynamic functions
at zero temperature.
However, increasing the pressure (and density), the multiparticle averages are reduced to
the product of single-particle averages which determine the dependence on thermodynamic
parameters. This enables to relate the thermodynamic functions by relations for an ideal gas,
where the mean value of quantum fluctuations is regarded as the effective temperature
of the inhomogeneous system.

Note that a similar splitting of many-particle correlators into single-particle functions
is observed in the model with pairwise interaction in the Thomas-Fermi approximation.
This emphasizes the importance of the model with three-particle interaction along with
explicit accounting of quantum fluctuations.

We associate the revealed changes in properties with the first-order
phase transition described in our previous work~\cite{GN21}. Besides, as found in
Ref.~\cite{GN21}, a description of rotational curves using this model is realized in
the vicinity of the smallest value of the chemical potential (the ground state) and
at a short pairwise scattering length $a$, the magnitude of which we reproduced above.
Then, we can compare our expression (\ref{ED}) for the internal energy with
the well-known one, where the term of three-particle interaction arises due to
quantum corrections in a system with pairwise interaction and the use of renormalization
group calculations~\cite{BrN99,LY57,Wu59,BBR}: 

\begin{eqnarray}
&&{\cal E}=\frac{{\cal Q}(n)}{2}n^2+\frac{{\cal B}(n)}{3}n^3,\label{Eren}\\
&&{\cal Q}(n)=\frac{4\pi\hbar^2 a}{m}\left(1+\frac{128}{15\sqrt{\pi}}\sqrt{na^3}\right),\\
&&{\cal B}(n)=\frac{16\pi\hbar^2 a^4}{m}\left(4\pi-3\sqrt{3}\right)\left[\ln{\left(na^3\right)}+2d\right].
\end{eqnarray}
Here $n$ is the particle density of homogeneous system of bosons, $a$ is their
pairwise $s$-scattering length, $d$ is a numeric parameter.

Note that the internal energy ${\cal E}$ is proportional to inverse of the mass of
(dark matte) particle.

However, we also admit that the origin of the three-body interaction can be explained
by relativistic effects~\cite{Chav2018,Chav2021} or by the very nature of dark matter
particles.

It is interesting to note the attempts to generalize the expression (\ref{Eren})
in the case of the unitary regime with an infinitely long scattering length $a$
\cite{C02,CW11}.
This is motivated, in particular, by the study of Bose condensates of alkali atoms,
the strong interaction between which can cause the molecules synthesis,
recombination, and three-particle losses. Hypothetically, similar processes can
result in multicomponent structure of dark matter , if the resulting composites are
stable enough and have a short scattering length. Although, due to the greater mass,
the Compton wavelength of the composite would already become shorter than that of
individual constituents.

Within the framework of an efficient model of chemical kinetics in Section~4,
we show that a long scattering length, but not a high density, is necessary for
the formation of few-particle complexes. Even without knowledge of the nature
of the bonds, it is clear that the appearance of dimers and trimers therein
cannot be compared with the synthesis of molecules from atoms that are complicated
by themselves. Moreover, there is no evidence that the processes under
consideration are similar to the production of baryons: quark-antiquark mesons and
three-quark nucleons or hyperons.

We assume that the Efimov effect could be appropriate mechanism of forming
the trimers from dark bosons. The existence of such an effect is successfully
confirmed by experiments with cold-atom systems~\cite{FG10} and is investigated~\cite{Mus21}.
Comparison of manifestations of this effect in the two different situations shows
a number of similar features.

To simulate unstable composites consisting of no more than three particles,
we consider a random walk model in Section~3. By adjusting the coupling
constants, it is able to obtain configurations with the dominance of composites
over long time intervals (long oscillation period). This model is quite useful
to visualize the decay of correlators in the Bose condensate with a three-body
interaction.

We expect that detailed quantum analysis may provide results applicable to the description
of dark matter, which will be published elsewhere.
 
\vspace{6pt}

\section*{Acknowledgments}

{Both authors acknowledge support from the National Academy of
Sciences of Ukraine by its priority project ``Fundamental properties of the
matter and its manifestation in the micro world, astrophysics and
cosmology in relativistic collisions of nuclei and in the early Universe''.}

\end{document}